\documentclass[sigconf,screen]{acmart}

\usepackage{array}
\usepackage{graphicx} 
\usepackage[skip=1pt]{caption}
\usepackage{subcaption} 
\usepackage[inline]{enumitem}

\usepackage{bbm}
\usepackage{nameref,hyperref}
\PassOptionsToPackage{table}{xcolor}

\usepackage{ulem}

\usepackage{csquotes}

\usepackage{listings}

\usepackage{acronym}

\AtBeginDocument{%
  \providecommand\BibTeX{{%
    \normalfont B\kern-0.5em{\scshape i\kern-0.25em b}\kern-0.8em\TeX}}}

\acrodef{GNN}{graph neural network}
\acrodef{LBSN}{location-based social network}
\acrodef{LLM}{large language model}
\acrodef{POI}{point-of-interest}
\acrodef{RNN}{recurrent neural network}
\acrodef{PEFT}{Parameter-Efficient-Fine-Tuning}
\acrodef{NF}{NormalFloat}
\acrodef{FP}{Floating Point}
\acrodef{BF}{BrainFloating}
\acrodef{LoRA}{Low-Rank Adaptation}
\newcommand{\headernodot}[1]{\vspace*{1.5mm}\noindent\textbf{#1}}
\newcommand{\header}[1]{\headernodot{#1.}}

\copyrightyear{2024} \acmYear{2024} \setcopyright{acmlicensed}\acmConference[SIGIR '24]{Proceedings of the 47th International ACM SIGIR Conference on Research and Development in Information Retrieval}{July 14--18, 2024}{Washington, DC, USA} \acmBooktitle{Proceedings of the 47th International ACM SIGIR Conference on Research and Development in Information Retrieval (SIGIR '24), July 14--18, 2024, Washington, DC, USA} \acmDOI{10.1145/3626772.3657840} \acmISBN{979-8-4007-0431-4/24/07}

\author{Peibo Li}
\orcid{0009-0001-0201-3564} 
\affiliation{%
  \institution{University of New South Wales}
  \city{Sydney}
  \country{Australia}
}
\email{peibo.li@student.unsw.edu.au}

\author{Maarten de Rijke}
\orcid{0000-0002-1086-0202} 
\affiliation{%
  \institution{University of Amsterdam}
  \city{Amsterdam}
  \country{The Netherlands}
}
\email{m.derijke@uva.nl}

\author{Hao Xue}
\orcid{0000-0003-1700-9215} 
\affiliation{%
  \institution{University of New South Wales}
  \city{Sydney}
  \country{Australia}
}
\email{hao.xue1@unsw.edu.au}

\author{Shuang Ao}
\orcid{0009-0008-3321-8416}
\affiliation{%
  \institution{University of New South Wales}
  \city{Sydney}
  \country{Australia}
}
\email{shuang.ao@unsw.edu.au}

\author{Yang Song}
\orcid{0000-0003-1283-1672}
\affiliation{%
  \institution{University of New South Wales}
  \city{Sydney}
  \country{Australia}
}
\email{yang.song1@unsw.edu.au}

\author{Flora D. Salim}
\orcid{0000-0002-1237-1664}
\affiliation{%
  \institution{University of New South Wales}
  \city{Sydney}
  \country{Australia}
}
\email{flora.salim@unsw.edu.au}

\renewcommand{\shortauthors}{Peibo Li et al.}



\title[Large Language Models for Next Point-of-Interest Recommendation]{\texorpdfstring{Large Language Models\\ for Next Point-of-Interest Recommendation}{Large Language Models for Next Point-of-Interest Recommendation}}

\begin{document}

\begin{abstract}
The next \ac{POI} recommendation task is to predict users' immediate next \ac{POI} visit given their historical data. 
\Acl{LBSN} data, which is often used for the next \ac{POI} recommendation task, comes with challenges. 
One frequently disregarded challenge is how to effectively use the abundant contextual information present in \acl{LBSN} data. 
Previous methods are limited by their numerical nature and fail to address this challenge. 
In this paper, we propose a framework that uses pretrained \aclp{LLM} to tackle this challenge. 
Our framework allows us to preserve heterogeneous \acl{LBSN} data in its original format, hence avoiding the loss of contextual information. 
Furthermore, our framework is capable of comprehending the inherent meaning of contextual information due to the inclusion of commonsense knowledge. 
In experiments, we test our framework on three real-world \acl{LBSN} datasets. 
Our results show that the proposed framework outperforms the state-of-the-art models in all three datasets. Our analysis demonstrates the effectiveness of the proposed framework in using contextual information as well as alleviating the commonly encountered cold-start and short trajectory problems.
Our source code is available at: https://github.com/neolifer/LLM4POI
\end{abstract}

\maketitle

\acresetall

\section{Introduction}


\Acp{LBSN} have experienced massive growth, capitalizing on developments in mobile and localization techniques, as they provide rich location-based geo-information. 
Next \Ac{POI} recommendation, as one of the applications that use \ac{LBSN} data, predicts users' next \ac{POI} visit, given their historical trajectories. Existing next \ac{POI} methods \cite{lim_stpudgat_2020,zhang2022next,yang2022getnext,yan2023spatio} focus on the short trajectory and cold-start problem, where users with a small amount of data and short trajectories are harder to predict. While these methods alleviate the short trajectory and cold-start problem, they do not fully explore the potential of \ac{LBSN} data.
In particular, the rich contextual information contained in \ac{LBSN} data has the potential to precisely model users' behavior. By using contextual information, we can understand the data in a way beyond statistics and even derive patterns that do not exist in the data. For example, the data showing a user who frequently visits college buildings during teaching periods could indicate the user's identity as a student or a college staff. Therefore, it is likely that this user will behave differently during vacation between teaching periods. 

To exploit such contextual information in \ac{LBSN} data, there are some substantial \textbf{challenges}: 
\begin{enumerate*}[label=(\textbf{\roman*)}]
    \item How to extract the contextual information from the raw data? And
    \item How to connect contextual information with commonsense knowledge to effectively benefit next POI recommendation?
\end{enumerate*}
Here, we consider \textit{contextual information} as time, POI category, and geo-coordinates. 
And we define \textit{commonsense knowledge} in the context of the next POI recommendation as the capability to understand the semantics of contextual information without additional data, and to connect certain joint patterns of contextual information with behaviors in the real world.
Existing next \ac{POI} recommendation methods \cite{lim_stpudgat_2020,zhang2022next,yang2022getnext,yan2023spatio} have two important \textbf{limitations} when dealing with contextual information:
\begin{enumerate*}[label=(\textbf{\roman*)}]
    \item Due to their numerical nature, they have to transform heterogeneous \ac{LBSN} data into numbers. For example, POI categories are often encoded from text into IDs. This transformation can result in a loss of inherent meaning associated with contextual information. 
    \item They rely exclusively on statistics and human designs to understand contextual information and lack an understanding of the semantic concepts provided by the contextual information.
\end{enumerate*}


\Acp{LLM} have demonstrated capabilities in a variety of tasks. 
Question-answering, in particular, has benefited from the commonsense knowledge embedded in \acp{LLM} \cite{zhao2023survey, NEURIPS2020_1457c0d6}. \acp{LLM} have a basic grasp of the concepts in daily life and can respond to users' questions using these concepts.
Inspired by this and the textual nature of \ac{LBSN} data, leveraging \acp{LLM} for the next \ac{POI} recommendation task seems a natural step. 
In our work, we adopt the pretraining and fine-tuning paradigm, and fine-tune pretrained \acp{LLM} on \ac{LBSN} data. 
As we will see below, by doing so, we are able to use a single \ac{LLM} to deal with all types of \ac{LBSN} data and better use contextual information.

\begin{figure}[t]
  \centering
  \begin{tabular}{@{}c@{}}
  \includegraphics[width=.9\linewidth]{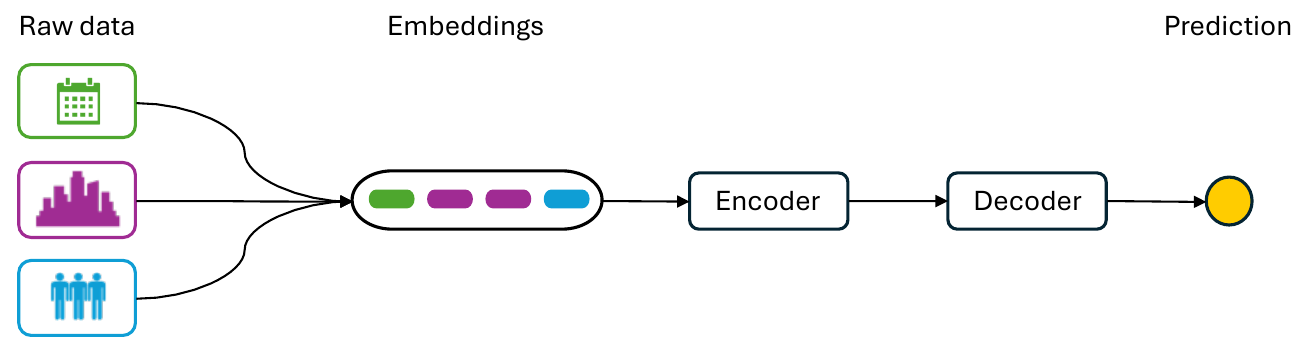} 
  \\
  \textbf{(a) Feature transformation paradigm.}
  \\
  \includegraphics[width=.9\linewidth]{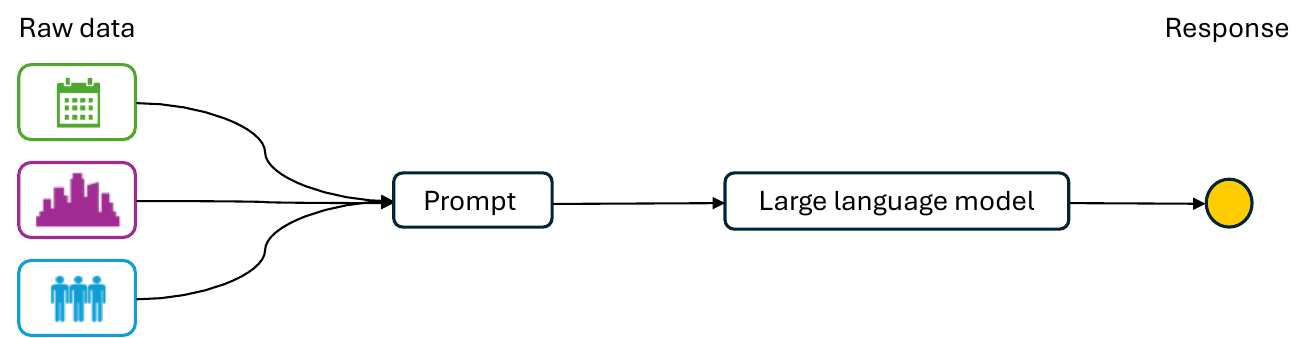} 
  \\
  \textbf{(b) \Acl{LLM}-based paradigm.}
  \end{tabular}
  \caption{Comparison of two paradigms for the next \ac{POI} task: (a) a typical feature transformation paradigm and (b) the proposed language model-based paradigm.}
  \label{fig:paradigms}
\end{figure}

More specifically, to address \textbf{challenge~(i)}, we transform the next \ac{POI} task into a question-answering task. 
We convert text-formed raw data into prompts constructed by blocks of sentences. 
Each block serves as a different module, and each sentence in it contains the necessary information for that module in its original format. 
Therefore, all heterogeneous data can be fed into a single model with tokenization that still keeps the contextual information. 
As illustrated in Figure~\ref{fig:paradigms}, unlike existing methods that we categorize as feature transformation paradigms, where data needs to be transformed and fed into different embedding layers, our method allows the data to be directly used in its original format. 
We also propose a notion of trajectory similarity based on prompts, which is used for the cold-start problem.

For \textbf{challenge ~(ii)}, we use pretrained \acp{LLM} that have been trained on a large corpus with rich commonsense knowledge. 
The contextual information in the tokenized data can be understood with its inherent meaning rather than being treated just as a code. 
As an example, in Table~\ref{fig:category}, we present the \ac{POI} category names in a real-world dataset, categorized by their context, as done by ChatGPT.\footnote{We provide ChatGPT with the dataset file and ask it to find the unique POI category names. Then we use the prompt ``Can you list the category names that have intersections by their context?" to get the content in Table \ref{fig:category}. \url{https://chat.openai.com}} 
This demonstrates that \acp{LLM} are capable of understanding the inherent meaning of contextual information in \ac{LBSN} data.

\begin{table}[h]
\caption{\ac{POI} categories in a real-world dataset summarized by ChatGPT. The left column is summarized by ChatGPT. We first feed the raw csv file to ChatGPT and ask it to find the unique POI category names. Then we use the prompt `Can you list the category names that have intersections by their context?' to generate the categories that are summarized from the right column, which represents the subcategories from the raw data.}
\label{fig:category}
\centering
\begin{tabular}{ll}
\toprule
\textbf{Category (GPT)} & \textbf{Subcategories/Category Names (Raw Data)} \\
\midrule
Food and Dining & \begin{tabular}[c]{@{}l@{}}Restaurant (General)\\ \textbf{Specific Restaurants}: American, Asian, \\
Italian, Mexican, Korean, Thai, \\
Mediterranean, Caribbean\\ \textbf{Specific Food Types}: Seafood Restaurant,\\BBQ Joint, Steakhouse, 
Pizza Place, \\ \textbf{Other Dining}: Café, Bistro, Diner, Bakery, \\
Food Truck,  Deli / Bodega, Dessert Shop\end{tabular} \\
\midrule
Beverages & \begin{tabular}[c]{@{}l@{}} Bar, Beer Garden, Coffee Shop,\\ Brewery, Tea Room, Juice Bar\end{tabular} \\
\midrule
Accommodations & \begin{tabular}[c]{@{}l@{}} Hotel, Motel, Hostel, Bed and Breakfast\end{tabular} \\
\midrule
\begin{tabular}[c]{@{}l@{}}Shopping and \\Retail \end{tabular}& \begin{tabular}[c]{@{}l@{}} Department Store, Clothing Store,\\ Electronics Store, Bookstore,\\ Market, Mall, Miscellaneous Shop\end{tabular} \\
\midrule
\begin{tabular}[c]{@{}l@{}}Outdoor and \\Recreation \end{tabular}& \begin{tabular}[c]{@{}l@{}} Park, Beach, Zoo, Garden, Plaza, \\Other Great Outdoors,\\ Playground, Campground\end{tabular} \\
\midrule
\begin{tabular}[c]{@{}l@{}}Arts and \\Entertainment\end{tabular} & \begin{tabular}[c]{@{}l@{}} Museum, Art Museum, Theater, Cinema, \\ Concert Hall, Music Venue, Art Gallery, \\ Comedy Club, Performing Arts Venue\end{tabular} \\
\midrule
\begin{tabular}[c]{@{}l@{}}Health and \\Fitness \end{tabular}& \begin{tabular}[c]{@{}l@{}} Gym / Fitness Center, Spa / Massage,\\ Medical Center, Yoga Studio\end{tabular} \\
\midrule
\begin{tabular}[c]{@{}l@{}}Travel and \\Transport \end{tabular}& \begin{tabular}[c]{@{}l@{}} Airport, Train Station, Bus Station,\\ Subway, Ferry, Taxi\end{tabular} \\
\midrule
\begin{tabular}[c]{@{}l@{}}Educational \\Institutions \end{tabular}& \begin{tabular}[c]{@{}l@{}} School, University, Library, Museum, \\ College Academic Building\\\end{tabular} \\
\midrule
\begin{tabular}[c]{@{}l@{}}Professional and \\Office \end{tabular}& \begin{tabular}[c]{@{}l@{}} Office, Corporate Building,\\ Conference Room\end{tabular} \\
\midrule
Residential & \begin{tabular}[c]{@{}l@{}} Residential Building (Apartment / Condo),\\ Home (private), Housing Development\end{tabular} \\
\midrule
\begin{tabular}[c]{@{}l@{}}Cultural and \\Religious \end{tabular}& \begin{tabular}[c]{@{}l@{}} Church, Temple, Shrine, Synagogue\end{tabular} \\
\bottomrule
\end{tabular}
\end{table}

\header{Contributions}
The main contributions of our work are as follows:
\begin{enumerate}[leftmargin=*]
    \item We propose LLM4POI, a framework to use pretrained \aclp{LLM} for the next \ac{POI} recommendation task, which brings commonsense knowledge for making use of the rich contextual information in the data. To the best of our knowledge, we are the first to fine-tune language models on standard-sized datasets to exploit commonsense knowledge for the next \ac{POI} recommendation task.
    \item We propose a prompt-based trajectory similarity to combine information from historical trajectories and trajectories from different users. With that, our proposed recommendation model is able to alleviate the cold-start problem and make predictions with improved accuracy over trajectories of various lengths.
    \item We conduct an extensive experimental evaluation on three real-world \acp{LBSN} datasets, which shows that our proposed next \ac{POI} recommendation model substantially outperforms state-of-the-art next \ac{POI} recommendation models in all three datasets.
\end{enumerate}

\section{Related Work}
\begin{figure*}
    \centering
    \includegraphics[width=0.85\textwidth]{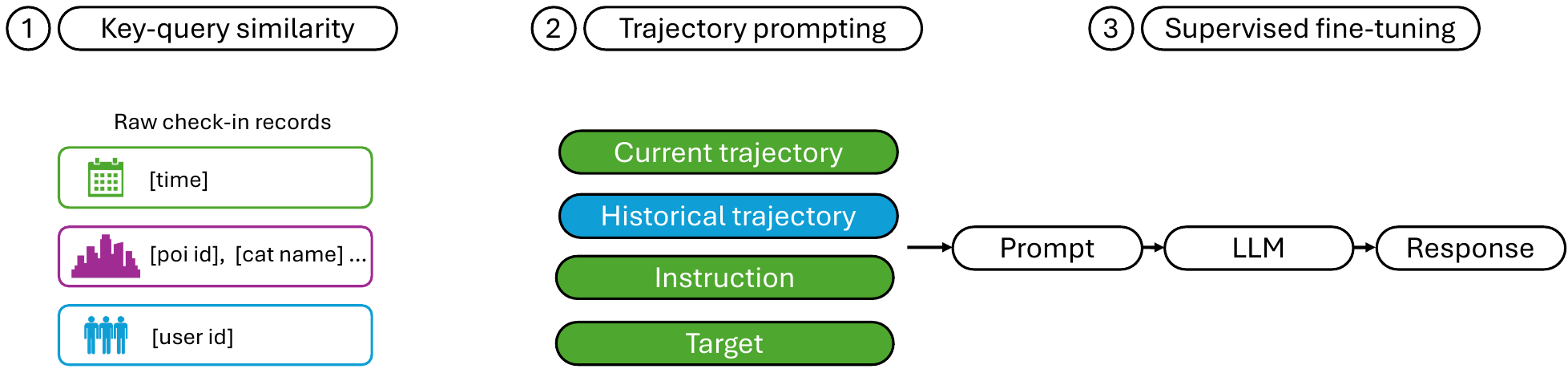}
    \caption{Our overall \acl{LLM}-based framework for next \ac{POI} recommendation.}
    \label{framework}
\end{figure*}

\subsection{Next \ac{POI} Recommendation}
\textbf{Sequence-based models.} 
Early work on next \ac{POI} recommendation often treated the next POI recommendation as a sequential recommendation task. 
Therefore, methods that had been widely used for other sequential recommendation tasks were adapted. 
For instance, the next POI recommendation task was first introduced by \citet{cheng2013you}, and they adapted FMPC \cite{rendle2010factorizing}, implementing a localized region constraint where only neighborhood locations are considered for each user. 
\citet{he_inferring_2016} combined the personalized Markov chain with the latent pattern by incorporating the softmax function. 
However, these methods are less capable of capturing complex sequential patterns compared to deep neural networks. 

Subsequent work has begun to apply RNN-based models with the rise of deep learning. \citet{kong_hst-lstm_2018} proposed HST-LSTM, where they added spatial-temporal factors into LSTM gates to guide the learning and further employed a hierarchical extension to model the periodicity of the visit sequence. 
LSTPM \cite{sun_where_2020} employed three LSTM modules, utilizing non-local neural operations and a short-term preference modeling module using geo-dilated LSTM. 
PLSPL \cite{wu2020personalized} combined an embedding layer with the attention mechanism to learn long-term preferences and leveraged two LSTM models to model short-term preferences at both the location and category levels. 
STAN \cite{luo2021stan} uses a multimodal embedding layer to learn the representation of user, location, time, and spatial-temporal effect, with a bi-layer attention architecture to learn the explicit spatial-temporal relevance within the trajectories. 
CFPRec \cite{zhang2022next} focused on the multi-step future plan of users by adopting an attention mechanism to extract future references from past and current preference encoders that are transformer and LSTM encoders. 
These sequence-based models often become confined to a local view and also suffer from short trajectories and the cold-start problem where inactive users have limited data. Our method deploys key-query similarity, which allows us to combine information from different users, alleviating the cold-start problem. 

\header{Graph-based models} 
More recently, graph-based methods have been incorporated to address the limitations of sequence-based models. 
STP-UDGAT \cite{lim_stpudgat_2020} were the first to use a graph attention network \cite{velivckovic2018graph}, enabling users to selectively learn from others in a global view. 
\citet{zhang2022next} proposed a hierarchical multi-task graph recurrent network (HMT-GRN) to learn user-POI and user-region distribution, employing a GRN to replace the LSTM unit to learn both sequential and global spatial-temporal relationships between POIs. 
DRGN \cite{wang2022learning} investigated the intrinsic characteristics of POIs by learning disentangled representation from both distance-based and transition-based relation graphs through a GCN layer. 
GETNEXT \cite{yang2022getnext} addressed the cold start problem by exploiting collaborative signals from other users and proposing a global trajectory flow map and a novel graph-enhanced transformer model. 
STHGCN \cite{yan2023spatio} alleviated cold-start issues by constructing a hypergraph to capture higher-order information, including user trajectories and collaborative relations.
Although graph-based models handle the cold-start problem, they are not capable of combining contextual information with commonsense knowledge. Our method avoids contextual information loss by trajectory prompting, and the \acp{LLM} that we use contains commonsense knowledge to understand contextual information.

\subsection{\acp{LLM} for Time-series Data}
\acp{LLM} have proven to be effective for time-series data. 
The study by SHIFT \cite{xue2022translating} approached human mobility forecasting as a language translation problem rather than a traditional time-series problem, utilizing a sequence-to-sequence language model complemented by a mobility auxiliary branch. 
AuxMobLCast \cite{xue2022translating} further investigated prompt engineering on time-series data. 
LLM4TS \cite{chang2023llm4ts} employs a two-stage fine-tuning approach, initially applying supervised fine-tuning to align the LLM with time-series data, followed by downstream task-specific fine-tuning. Inspired by these works, we design trajectory prompting specifically for \ac{LBSN} data, allowing us to transform the next \ac{POI} recommendation task into a question-answering task.
\subsection{\acp{LLM} for Recommender Systems}
Recently, many works have adopted \acp{LLM} on recommender systems. \citet{zhang2023prompt} designed multiple prompt templates for different perspectives of news data and did prompt-learning BERT \cite{devlin-etal-2019-bert} to produce binary answers to templates. Then multi-prompt ensembling was applied to get final predictions. \citet{harte2023leveraging} proposed three approaches to leverage \acp{LLM} for sequential recommendation. They first compute the embeddings of items and then make recommendations based on the similarity of item embeddings. They also directly fine-tune \acp{LLM} to do a prompt completion where the prompt contains a list of item names without the last item and \acp{LLM} are asked to complete the prompt with the name of the last product. They also enhanced existing sequential models with embeddings from \acp{LLM}. \citet{wang2023would} applied in-context learning with \acp{LLM} for next \ac{POI} recommendation. Our method not only fine-tune \acp{LLM} for the next \ac{POI} recommendation, but also has carefully designed task-specific trajectory similarity to further utilize the power of \ac{LLM}.

\section{Problem Definition}

The research problem that we address in this paper is to fine-tune \acp{LLM} for the task of next \ac{POI} recommendation. 
The problem can be formalized as follows. 
Consider a dataset \( \mathcal{D} \) of user check-in records. 
Each record is represented by a tuple \(q =  (u, p, c, t, g) \), where:
\begin{itemize}[leftmargin=*]
    \item \( u \) denotes a user from the set \( U = \{u_1, u_2, \ldots, u_N\} \), where \( N \) is the total number of users;
    \item \( p \) is a point of interest (POI) from the set $ P = \{p_1$, $p_2$, \ldots, $p_M\}$, where \( M \) is the number of distinct POIs;
    \item \( c \) specifies the category of the POI;
    \item \( t \) represents the timestamp of the check-in; and
    \item \( g \) signifies the geometric coordinate of the POI.
\end{itemize}

\noindent%
Given a time interval \( \Delta t \), trajectories for a user \( u \) are formed by splitting the check-in records based on this interval. 
Each trajectory \( T_i^u \) up to timestamp \( t \) for user \( u \) is given by:
\[ 
T_i^u(t) = \{(p_1, c_1, t_1, g_1), \ldots, (p_k, c_k, t_k, g_k)\}, 
\]
where \( t_1 < t_2 < \cdots < t_k = t \) and \( t_{k} - t_1 \leq \Delta t \).

Given this set of historical trajectories \( \mathcal{T}_u = \{T_1^u, T_2^u, \ldots, T_L^u\} \) for user \( u \), where \( L \) represents the number of trajectories for \( u \), the objective is to predict the POI \( p_{k+1} \)  for a new trajectory \( {T'}_i^u(t) \), where user \( u \) will check in at the immediate subsequent timestamp \( t_{k+1} \).

\section{Methodology}

The overall framework of our work is presented in Figure \ref{framework}. Our method includes three components: trajectory prompting, key-query similarity, and supervised fine-tuning for \acp{LLM}. First, the raw data is used to construct the prompt and compute the prompt-based key-query similarity. Trajectory prompting uses both raw data and the key-query similarity to form the prompts for the \ac{LLM}. The \ac{LLM} is then trained with supervised fine-tuning using the prompts.

\subsection{Trajectory Prompting}

Inspired by \cite{zhou2022learning, radford2021learning, xue2022leveraging}, we propose trajectory prompting to convert sequences of user check-in data into a natural language question-answering format for \acp{LLM} to follow the instruction from the prompt and generate the POI recommendation. This transformation is crucial in leveraging the power of pretrained \acp{LLM}. The idea of trajectory prompting is to unify heterogeneous \ac{LBSN} data into meaningful sentences that can be fed into \acp{LLM}. Specifically, we construct prompts by designing different blocks of sentences for their respective purposes. As shown in Table \ref{prompt}, a prompt consists of the current trajectory block, the historical trajectory block, the instruction block, and the target block.

\begin{table}[ht]
\caption{Structure of prompts and check-in record. Red indicates the current trajectory block. Purple indicates the historical trajectory block. Orange indicates the instruction block. Blue indicates the target block.}
\label{prompt}
\centering
\begin{tabular}{l m{0.3\textwidth}}
\toprule
\textbf{prompt} &
{\textless{}question\textgreater{}}
\textcolor{red}{The following is a trajectory of user [user id]: [check-in records]. }
\textcolor{purple}{There is also historical data: [check-in records]. }
\textcolor{orange}{Given the data, at [time], which \ac{POI} id will user [user id] visit? Note that \ac{POI} id is an integer in the range from 0 to [id range].}
\textcolor{blue}{\textless{}answer\textgreater{}: At [time], user [user id] will visit \ac{POI} id [poi id].} \\ 
\hline
\textbf{check-in record} &
At [time], user [user id] visited \ac{POI} id [poi id] which is a/an [poi category name] with category id [category id]. \\
\bottomrule
\end{tabular}
\end{table}

There are \textit{check-in record} sentences for both the current trajectory block and the historical trajectory block. These sentences contain the necessary information in a check-in record (e.g., user ID, timestamp, \ac{POI} category name, \ac{POI} category ID). Specifically, for each check-in record 
$q=(u,p,c,t,g)$, we form the sentence as ``At [time], user [user id] visited \ac{POI} id [poi id] which is a/an [poi category name] with category id [category id].'' 
We do not include geo-coordinates in the sentence to save the number of tokens and we also find that \acp{LLM}, without specifically fine-tuning on map data, are not able to distinguish geo-coordinates well. The check-in records then form trajectories. Note that for the current trajectory block, there will only be one trajectory from the current user, and there can be multiple trajectories from arbitrary users for the historical trajectory block.

The \textit{current trajectory block} provides information for the current trajectory, excluding the last entry. The \textit{historical trajectory block} incorporates historical information from both the current user and other users who have similar behavior patterns to the current user, which is used for dealing with short trajectory and cold start problems. The details of selecting historical trajectories will be explained in Section \ref{similarity}. The \textit{instruction block} guides the model on what to focus on and also reminds the model of the range of \ac{POI} IDs since the \ac{POI} IDs generated by \acp{LLM} are not by simple argmax over the output probabilities of \acp{LLM} that are for the entire vocabulary. The \textit{target block} contains the timestamp, user ID, and \ac{POI} ID for the check-in record to be predicted, which serves as the ground truth for fine-tuning and evaluation. The target block is excluded from the input during prediction. We have tested adding \ac{POI} category information in both the instruction block and target block, expecting to encourage the model to pay more attention to the relation between the \ac{POI} ID and the \ac{POI} category. However, it turned out that the performance did not show a significant difference, and the model might have already learned that inner relationship.

Our approach of prompting with blocks of sentences allows us to integrate the heterogeneous \ac{LBSN} data into meaningful sentences in its original format. The design of using blocks is easy to modify and extend.
\subsection{Key-Query Pair Similarity}

\label{similarity}
To capture patterns of users' behaviors from their historical trajectories and different users' trajectories, we propose a key-query pair similarity computation framework that suits trajectories in a natural language format. We treat each trajectory differently based on its respective position in the prompt. When a trajectory is considered for the current trajectory block, it is treated as the key, while any trajectory whose end time is earlier than this trajectory is treated as a query. We compute the similarity for all key-query pairs. Subsequently, we select queries with high similarity values for the historical trajectory block. This approach allows us to incorporate information from other users' trajectories that exhibit similar behavior patterns into the current trajectory.

\begin{figure}[t]
    \centering
    \includegraphics[width=\linewidth]{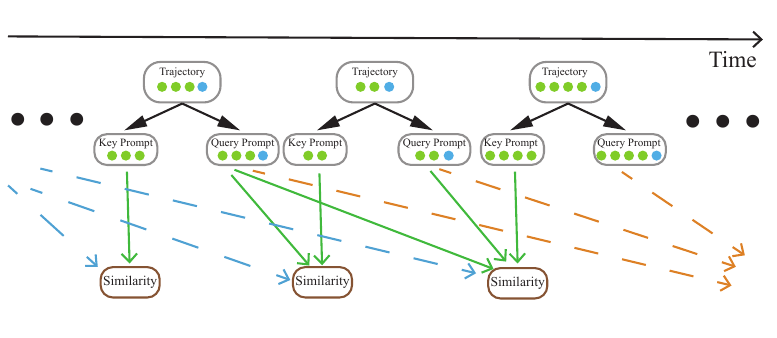}
    \caption{The process of forming and pairing key and query prompts. Each trajectory is made into a key prompt and a query prompt. The key prompt contains the check-in records excluding the last entry of the trajectory, while the query prompt contains the entire trajectory. A key prompt is paired with every query prompt representing the trajectories before the current trajectory.}
    \label{fig:phases-a}
\end{figure}

\begin{figure}[t]
    \centering
    \includegraphics[width=\linewidth]{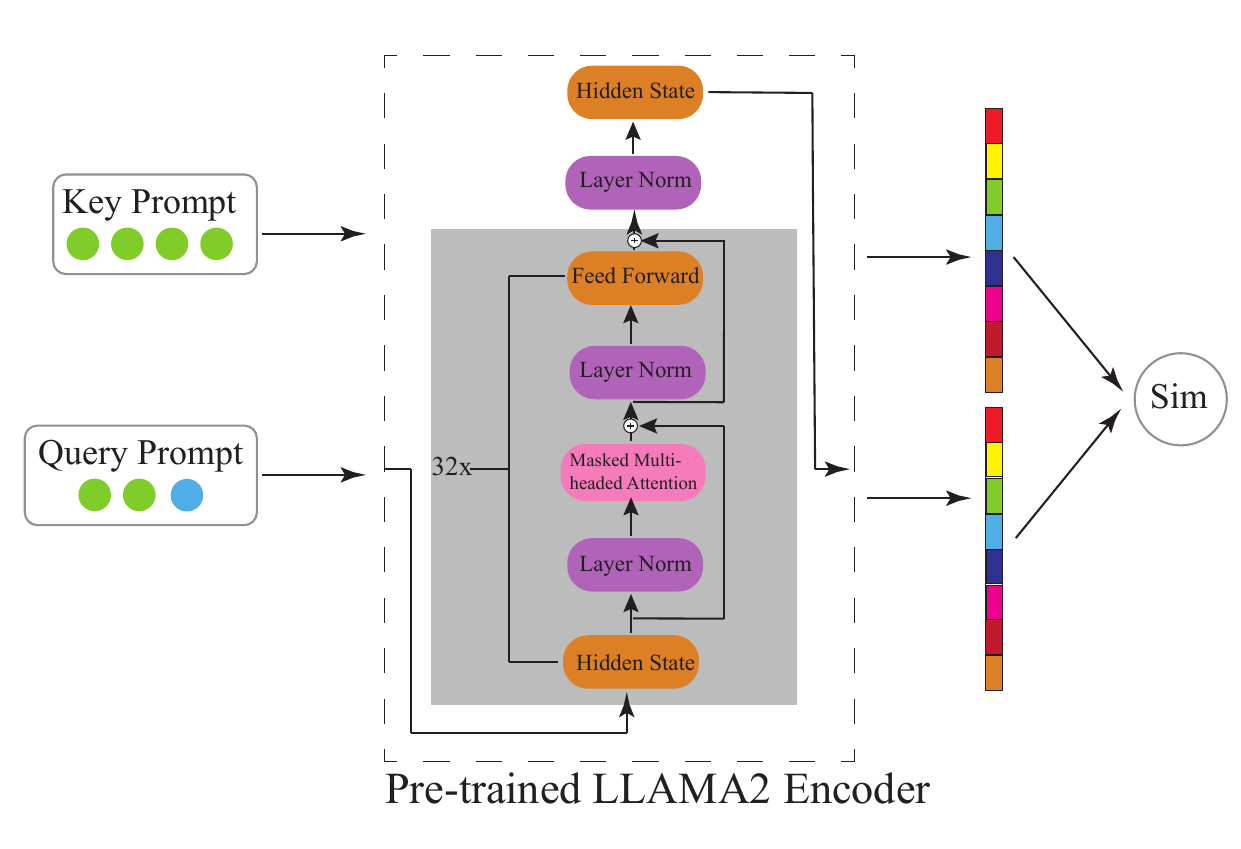}
    \caption{Similarity computation for each pair of key and query. Each pair of key and query prompts is fed into a pre-trained LLAMA2 separately. We use the last hidden layer embeddings to compute their cosine similarity.}
  \label{fig:phases-b}
\end{figure}

As illustrated in Figure \ref{fig:phases-a} and \ref{fig:phases-b} , we first form the key and query prompts for every trajectory. Specifically, we use the template for the current trajectory block. For the key prompts, we use the trajectories without their last entry, and for the query prompts, we use the entire trajectories. This is because when the query prompts are used as historical data, the historical trajectories are known, whereas the key prompts are treated as the current trajectories. For every key prompt, we compute the pairwise similarity for it and all the query prompts representing trajectories that have an end time earlier than the start time of the trajectories represented by the key prompt.

For each key and query prompt, we feed it into an \ac{LLM} encoder and get the embeddings from the last hidden layer. The process can be formulated as:
\begin{equation}
\begin{aligned}
h_{k_1} &= {\text{Transformer Block}_{(0)}}(\text{Tokenizer}(\text{Key})), \\
h_{k_l} &= {\text{Transformer Block}_{(l-1)}}(h_{k_{l-1}}), \\
E_k &= \text{LN}(h_{k_n}),
\end{aligned}
\end{equation}

\begin{equation}
\begin{aligned}
h_{q_1} &= {\text{Transformer Block}_{(0)}}(\text{Tokenizer}(\text{Query})), \\
h_{q_l} &= {\text{Transformer Block}_{(l-1)}}(h_{k_{l-1}}), \\
E_q &= \text{LN}(h_{k_n}),
\end{aligned}
\end{equation}

\noindent where \text{Tokenizer} is used for converting the prompt into a sequence of tokens, ${\text{Transformer Block}_{(i)}}$ represents the $i$-th Transformer block in the model, LN is the layer normalization, $E_k$ and $E_q$ are the final embeddings of the key and query, respectively.

After getting embeddings for every key and query, we compute the cosine similarity for each key-query pair. Formally, 
\begin{equation}
\text{Sim}(\mathbf{E_k}, \mathbf{E_q}) = \frac{\mathbf{E_k} \cdot \mathbf{E_q}}{\|\mathbf{E_k}\| \|\mathbf{E_q}\|}. 
\end{equation}

For each key, we select the top-$k$ queries with the highest similarity to the key. The trajectories represented by these queries are then used in the historical trajectory block for the key trajectory. The process can be expressed as
\begin{equation}
S(\text{key}) = \text{arg top}_k \{ \text{Sim}(q_i, \text{key}) \mid q_i \in Q \},
\end{equation}
where $Q$ is the set of queries with the end time earlier than the start time of the key.

\begin{figure}[t]
    \centering
    \includegraphics[width=\linewidth]{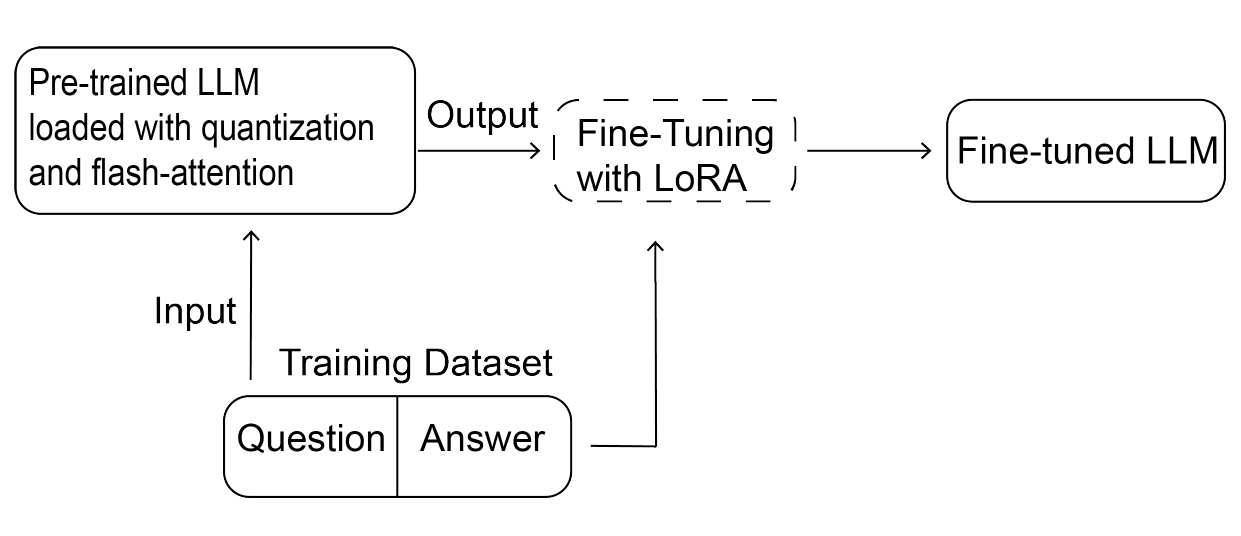}
    \caption{For data in the training set, the <question> part in the prompt is fed into the pre-trained LLM, 
and the <answer> part is used to guide the fine-tuning.}
  \label{fig:sft}
\end{figure}
\subsection{Supervised Fine-tuning}
Figure \ref{fig:sft} shows the overall our overall fine-tuning process.
Fine-tuning \acp{LLM} can be costly. We apply \Ac{PEFT} techniques during the fine-tuning stage.

\header{Low-rank adaptation} 
We apply LoRA \cite{hu2021lora}, freezing dense layers in LLMs and updating weights with rank decomposition matrices. For a pretrained weight matrix $W_0 \in R^{d\times k}$, we replace weight updates $W_0 + \Delta W$ with low-rank decomposition $W_0 + L_1L_2$, where $L_1\in R^{d\times r}$, $L_2\in R^{r\times k}$, and $r \ll min(d,k)$. During training, only $B$ and $A$ receive gradient updates. For $h=W_0x$, the modified forward pass is $h = W_0x+L_1L_2x$. LoRA is applied only to attention layers; MLP layers are frozen during fine-tuning. For an attention layer with 4096 elements, LoRA reduces trainable parameters to 0.78\% with a rank of 16, compared to full fine-tuning.

\header{Quantization} 
To reduce GPU memory usage, we apply quantization techniques \cite{dettmers2023qlora,dettmers20218}. Quantization involves converting high-bit data types into lower-bit ones. We use the 4-bit \ac{NF} proposed by \cite{dettmers2023qlora}, optimal for zero-mean normal distributions in [-1,1]. This process involves rescaling input tensors and applying quantization constants. To reduce memory overhead, double quantization is applied. NF4 is used for storage, while 16-bit \Ac{BF} is used for forward and backward passes. Note that the weight gradient is still only computed over the \Ac{LoRA} parameters.

\header{FlashAttention} 
Long trajectories and the historical trajectory block in the prompt require long context length from \acp{LLM}. A typical context length of 4096 is not enough for our purposes. Therefore, we apply FlashAttention-2 \cite{dao2022flashattention,dao2023flashattention}, which allows transformers to have long context lengths. 
\section{Experiment}

\subsection{Experimental Setup}
\subsubsection{Datasets}

We conduct experiments on three public datasets: Foursquare-NYC, Foursquare-TKY \cite{yang2014modeling}, and Gowala-CA \cite{cho2011friendship}. The first two datasets, collected over 11 months, comprise data from New York City and Tokyo, sourced from Foursquare. The Gowala-CA dataset, from the Gowalla platform, covers a broader geographical area and time period, encompassing California and Nevada. We utilize data that has been preprocessed as per the methods detailed by \citet{yan2023spatio}. The data is preprocessed as follows: 
\begin{enumerate*}[label=(\roman*)]
\item Filter out Points of Interest (\acp{POI}) with fewer than 10 visit records in history; 
\item Exclude users with fewer than 10 visit records in history; 
\item Divide user check-in records into several trajectories with 24-hour intervals, excluding trajectories that contain only one check-in record. 
\end{enumerate*}
The check-in records are also sorted chronologically: the first 80\% are used for the training set, the middle 10\% are defined as the validation set, and the last 10\% are defined as the test set. Note that the validation and test set has to contain all users and \acp{POI} that appear in the training set. 
The unseen users and \acp{POI} would be removed from the validation and test set.

\subsubsection{Baselines.}

We compare our model with the following baselines: 
\begin{itemize}[leftmargin=*]
    \item FPMC \cite{rendle2010factorizing}: rooted in the Bayesian Personalized Ranking framework, employs a typical Markov chain combined with matrix factorization to predict location transitions effectively.
    \item LSTM  \cite{hochreiter1997long}: A variant of RNN, is designed for processing sequential data. Unlike standard RNNs, LSTMs are capable of capturing both short-term and long-term dependencies in sequential patterns, making them more effective for a range of sequential data tasks.
    \item PRME \cite{feng2015personalized}: Utilizing a pairwise ranking metric embedding, this personalized ranking model effectively learns sequential transitions of \acp{POI} along with capturing user-\ac{POI} preferences in latent space.
    \item STGCN \cite{zhao2020spatio}: Based on LSTM, this model incorporates gating mechanisms to effectively model temporal and spatial intervals in check-in sequences, thereby capturing both short-term and long-term user preferences.
    \item PLSPL \cite{wu2020personalized}: This recurrent model employs an attention mechanism to learn short-term preferences and two parallel LSTM structures for long-term preferences, integrating both through a user-specific linear combination.
    \item STAN \cite{luo2021stan}: Leveraging a bi-layer attention architecture, STAN aggregates spatio-temporal correlations within user trajectories, learning patterns across both adjacent and non-adjacent locations as well as continuous and non-continuous visits.
    \item GETNext \cite{yang2022getnext}: A transformer-based approach, GETNext uses a global trajectory flow map that is user-agnostic to enhance next-\ac{POI} predictions, alongside proposing a GCN model for generating effective \ac{POI} embeddings.
    \item STHGCN \cite{yan2023spatio}: Constructing a hypergraph to capture inter and intra-user relations, STHGCN proposes a hypergraph transformer and solves the cold-start problem. 
\end{itemize}
    
\subsubsection{Our Models}
In our experiments we consider three versions of what we call ``our model'':
\begin{enumerate*}[label=(\roman*)]
\item LLM4POI*: Our model, using prompts only the current trajectory block without the historical trajectory block, where we use Llama-2-7b-longlora-32k \cite{longlora, touvron2023llama} as our base \ac{LLM}.
\item LLM4POI**: A variation on LLAMA2-7b where we use prompts with the historical trajectory block without applying key-query similarity, where only the historical trajectories from the current users are considered.
\item LLM4POI: A second variation on LLAMA2-7b where we use prompts with the historical trajectory block combined with key-query similarity, where historical trajectories from both the current users and other users are considered.
\end{enumerate*}
Below, unless stated otherwise, when we say ``our model,'' we refer to the LLM4POI variant.

\subsubsection{Evaluation Metrics}
For evaluation we regard the next \ac{POI} recommendation task as a top-1 recommendation problem.
The scenario we have in mind is one with ``extreme position bias''~\citep{farstein-2022-external}, where only a small amount of information can be presented to the user during a single interaction.
For example, a user who is on a business trip wishes to explore the new city before the meeting. In such a scenario, the user does not have the leisure to review multiple options. 
This is a scenario that is usually considered in next \ac{POI} recommendation, often as the primary scenario~\citep[see, e.g.,][]{yan2023spatio,yang2022getnext}.
Given this choice of scenario, our approach to the task as a type of question-answering problem is a natural fit.
We prioritize the delivery of the most pertinent and contextually suitable recommendation, mirroring the objective of providing a single, correct answer to a user's query.

The evaluation metric we use is Accuracy@1. 
It looks at what proportion of the test items would have been retrieved with the top-1 recommended list and can be formalized as:
\begin{equation}
    \text{Acc}@1 = \frac{1}{m} \sum_{i=1}^{m} \mathbbm{1}(\text{rank} \leq 1),
\end{equation}
where $\mathbbm{1}$ is the indicator function. Rank is the rank of the order of the correct prediction in the recommendation list. A larger value represents better performance.


\subsubsection{Implementation Details}
For fine-tuning, we use a constant learning rate schedule with a learning rate of $2 \times 10^5$, combined with a warm-up of 20 steps, a weight decay of 0, a batch size of 1 per GPU, and a sequence length of 32,768 tokens. For each dataset, we fine-tune the model for 3 epochs. We use approximately 300 historical check-in records to construct the historical trajectory block in the prompt. Our experiments are conducted on servers with Nvidia A100 GPUs.

\begin{table}[ht]
\centering
\caption{Performance comparison in terms of Acc@1 on three datasets. \checkmark and $\times$ in the History and Other Users columns indicate whether the model uses historical data or data from other users, respectively.}
\label{performance}
\begin{tabular}{lccccc}
\toprule
             Model   &History&Other & {NYC} & {TKY} & {CA} \\    \cline{4-6}
                &  &users &Acc@1 & Acc@1 & Acc@1  \\
\midrule
FPMC           &$\times$&$\times$ & 0.1003& 0.0814&  0.0383   \\
LSTM           &$\times$ &$\times$ & 0.1305& 0.1335&0.0665\\
PRME           &$\times$ &$\times$ & 0.1159 & 0.1052 & 0.0521   \\
STGCN           & $\times$ &$\times$ & 0.1799 & 0.1716 & 0.0961   \\
PLSPL          &$\times$ &$\times$ & 0.1917 & 0.1889 & 0.1072   \\
STAN            &$\times$ &$\times$ & 0.2231 & 0.1963 & 0.1104    \\
GETNext        & \checkmark & \checkmark & 0.2435 & 0.2254 & 0.1357  \\
STHGCN         &\checkmark & \checkmark & 0.2734 & 0.2950 & 0.1730  \\
\hline

LLM4POI*&$\times$& $\times$& 0.2356 &0.1517  & 0.1016  \\
LLM4POI**&\checkmark& $\times$& 0.3171 &0.2836  & 0.1683  \\
LLM4POI&\checkmark & \checkmark &\textbf{0.3372}&\textbf{0.3035}&\textbf{0.2065} \\
\bottomrule
\end{tabular}
\end{table}

\subsection{Main Results}
We compare the performance of our models and the baselines on three datasets, as shown in Table~\ref{performance}.

Our model substantially outperforms all baselines. Specifically, we observe improvements of 23.3\%, 2.8\%, and 19.3\% in top-1 accuracy on the NYC, TKY, and CA datasets, respectively, compared to the state-of-the-art STHGCN. Models utilizing historical data perform better than those that do not, and those incorporating data from other users see further performance boosts, highlighting the significance of short trajectory and cold-start problems in next \ac{POI} recommendation tasks. All models except STHGCN perform best in NYC, with noticeable performance drops in TKY and CA. NYC has the smallest number of users and \acp{POI} but a larger number of \ac{POI} categories than the other datasets, suggesting it has the easiest data to learn. In contrast, CA covers a much wider area than NYC and TKY, leading to data scarcity and significantly lower model performance.

\subsection{Analysis}
\subsubsection{User Cold-start Analysis}
\label{sec:cold_start}

Our approach incorporates the historical trajectory block and key-query trajectory similarity to tackle the cold-start problem by leveraging knowledge from diverse users. User activity status greatly impacts model performance, with active users providing more historical data and generally easier behavior patterns to learn. To assess our method's effectiveness with inactive users, we categorize users into inactive, normal, and very active groups based on the number of trajectories in the training set, designating the top 30\% users ranked by their number of trajectories as very active and the bottom 30\% as inactive.

We compare our model with STHGCN, which is designed to address the cold-start problem and has shown effectiveness with inactive users. The comparison, shown in Table \ref{active user}, reveals our model significantly improves performance for inactive users, more than doubling the baseline in NYC and increasing by over half in TKY and CA. This improvement underscores our method's ability to leverage information from similar users effectively, especially for those with limited historical data.

\begin{table}[ht]
\caption{User cold-start analysis on the NYC, TKY, and CA datasets.}
\label{active user}
\centering
\begin{tabular}{llccccc}
\toprule
User groups & Model & \multicolumn{1}{c}{NYC} & \multicolumn{1}{c}{TKY} & \multicolumn{1}{c}{CA}\\ \cline{3-5} 
                     &                & Acc@1     & Acc@1& Acc@1 \\ \midrule
Inactive             & STHGCN        & 0.1460              & 0.2164&0.1117\\
Normal               & STHGCN        & 0.3050              & 0.2659& 0.1620 \\
Very active          & STHGCN        & 0.3085         & 0.3464 &  0.2040\\ 
\midrule
Inactive             & LLM4POI           & 0.3417  & 0.3478  &   0.2132     \\
Normal               & LLM4POI           & 0.3841   & 0.3516 & 0.2057 \\
Very active          & LLM4POI           & 0.3088           & 0.2727 & 0.1920        \\ \bottomrule
\end{tabular}
\end{table}

\noindent%
Interestingly, our model performs better for inactive users than very active ones, contrasting the baselines' better performance with very active users. This suggests that the similar trajectories we identify for very active users are often from their own, leading to less diverse behavior patterns and less effective prediction of difficult data points. The less significant improvement in TKY and CA compared to NYC might be due to the higher user count in these datasets, limiting the collaborative information our method can utilize within the model's context length constraints.

\subsubsection{Trajectory Length Analysis}

The varying lengths of trajectories in the next \ac{POI} recommendation task, reflecting different user behaviors, pose another significant challenge. Short trajectories, with their limited spatio-temporal information and non-significant patterns, are particularly challenging, especially those with only one or two check-ins. While long trajectories contain more information, extracting useful patterns from them is also difficult. Our method's effectiveness varies with the trajectory length due to the context length limit, allowing fewer historical trajectories to be added for long trajectories compared to short ones. To explore the trade-off between long and short trajectories, we rank the lengths of trajectories in the test set, defining the top 30\% as long trajectories and the bottom 30\% as short trajectories, with the rest classified as middle trajectories.

Comparing our method to STHGCN, Table \ref{traj length} highlights our substantial improvement for short trajectories in NYC. We achieve an improvement of 24.4\% in top-1 accuracy for short trajectories and 31.6\% for middle trajectories in NYC, demonstrating our method's strong capability to integrate historical data for short trajectories. Interestingly, while we perform better for short trajectories in NYC, the opposite is true in TKY and CA. However, our model's performance does not vary significantly across different trajectory lengths, indicating a balanced trade-off between long and short trajectories.

\begin{table}[ht]
\centering
\caption{Trajectory length analysis on the NYC, TKY, and CA datasets.}
\label{traj length}
\begin{tabular}{llccccc}
\toprule
Trajectory types & Model & \multicolumn{1}{c}{NYC} & \multicolumn{1}{c}{TKY}  & \multicolumn{1}{c}{CA}\\ \cline{3-5} 
                     &                & Acc@1     & Acc@1& Acc@1 \\ \hline
Short             & STHGCN        & 0.2703              & 0.2787&0.1727\\
Middle               & STHGCN        & 0.2545              & 0.2823&0.1785  \\
Long          & STHGCN        & 0.3184         & 0.3116 & 0.1742 \\ \midrule
Short             & LLM4POI           & 0.3364  & 0.2876  &   0.1955    \\
Middle               & LLM4POI           & 0.3350   & 0.3013 &  0.1998\\
Long          & LLM4POI           & 0.3271         & 0.3083   &   0.2037    \\ \bottomrule
\end{tabular}
\end{table}

\subsubsection{Number of Historical Data Variants}

An important factor of our method is the number of historical trajectories used for the historical trajectory block. Since there is a token limit for the prompt, we cannot use all the historical trajectories but only the ones with the highest similarity to the current trajectory. Because trajectories vary in length, we use the number of check-in records in the historical trajectory block to evaluate the effect of the number of historical trajectories on performance. 

We compare our model trained and evaluated with different numbers of historical check-in records. As shown in Table \ref{num_hist}, we observe that the model archives the best performance in NYC with 100 historical check-in records and decreases as the number of check-in records grows. On the other hand, the model has the best performance in TKY with 300 historical check-in records, and the performance is positively correlated to the number of historical check-in records. Note that the performance in TKY only improves by a tiny margin when the number of historical check-in records increases from 200 to 300. The model performance remains closely in CA, with different numbers of different historical check-in records. The results indicate that using more historical data does not necessarily improve the model performance. We can use less historical data to reduce the token size of prompts but still achieve competitive model performance, which also speeds up the training and inference.

\begin{table}[ht]
\caption{Analysis on NYC, TKY, and CA dataset for LLM4POI trained on prompts with different numbers of historical trajectories.}
\label{num_hist}
\centering
\begin{tabular}{cccc}
\toprule
\multicolumn{1}{c}{Number of historical}  &  \multicolumn{1}{c}{NYC} & \multicolumn{1}{c}{TKY} & \multicolumn{1}{c}{CA}\\ \cline{2-4} 
 check-ins&                Acc@1     & Acc@1 & Acc@1\\ \midrule
100                     & 0.3420              & 0.2166 &0.2056\\
200                       & 0.3400              & 0.3023 &0.2035 \\
300                 & 0.3372         & 0.3035 & 0.2065 \\ 
\bottomrule
\end{tabular}
\end{table}

\subsubsection{Generalization to Unseen Data Analysis} \label{sec:unseen}
The fact that our approach does not rely on a linear classifier to output the \ac{POI} IDs but predicts with purely language modeling allows us to evaluate our models, fine-tuned on one dataset, on unseen data without any further training. We fine-tune our models on one of the NYC, TKY, and CA and then evaluate them on the rest.

\begin{figure}
    
    \includegraphics[width=\linewidth]{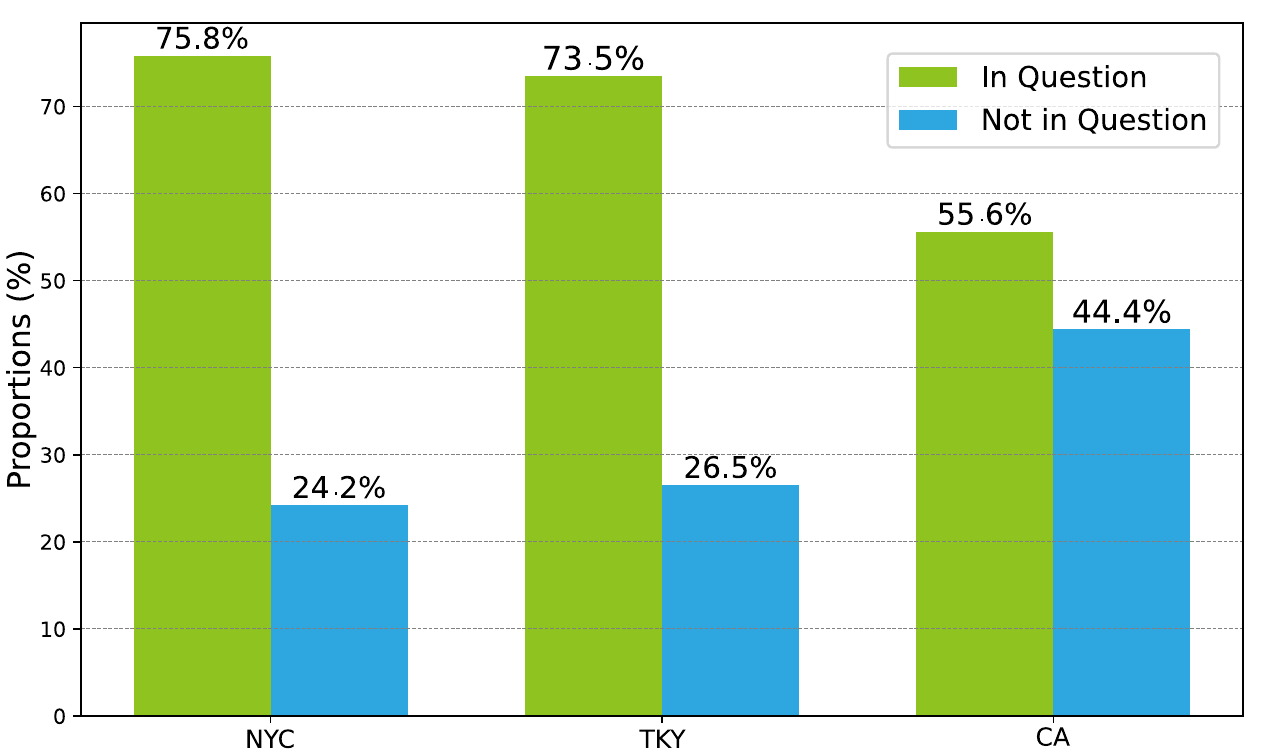}

    \caption{The proportion of test set prompts where the answer \ac{POI} IDs are included within the questions, in their respective datasets.}
    \label{fig:portion}
\end{figure}

\begin{figure}
    \includegraphics[width=\linewidth]{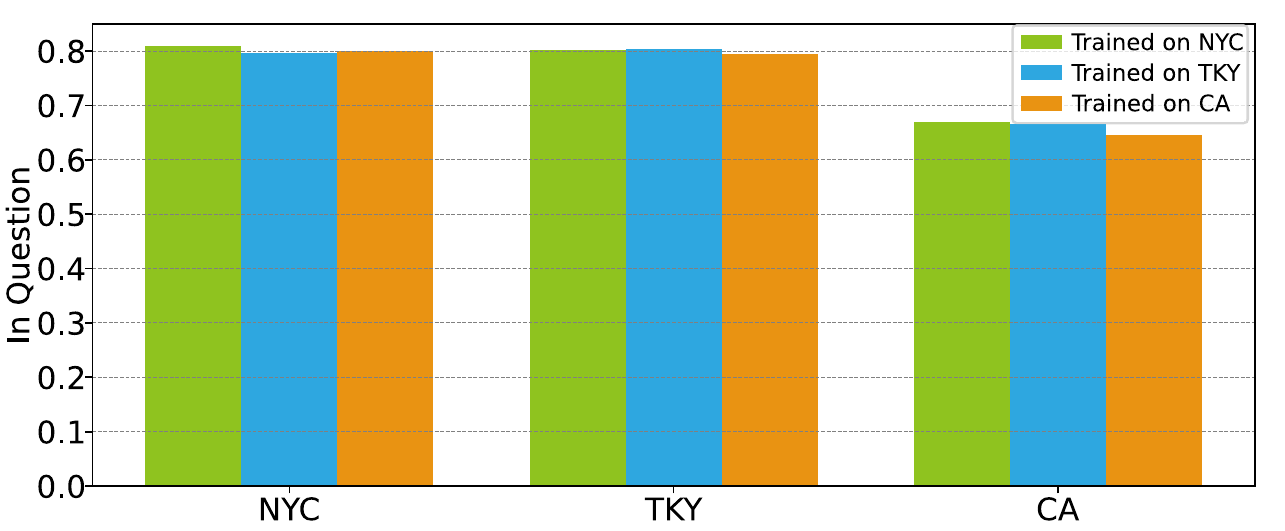}
    \caption{The proportion of test set prompts where the answer \ac{POI} IDs are included within the questions, in their respective datasets, for the correct predictions made by the models trained on the NYC, TKY, and CA.}
    \label{fig:fourth}
\end{figure}

\begin{figure*}[ht]

  \centering
    \includegraphics[width=0.92\linewidth]{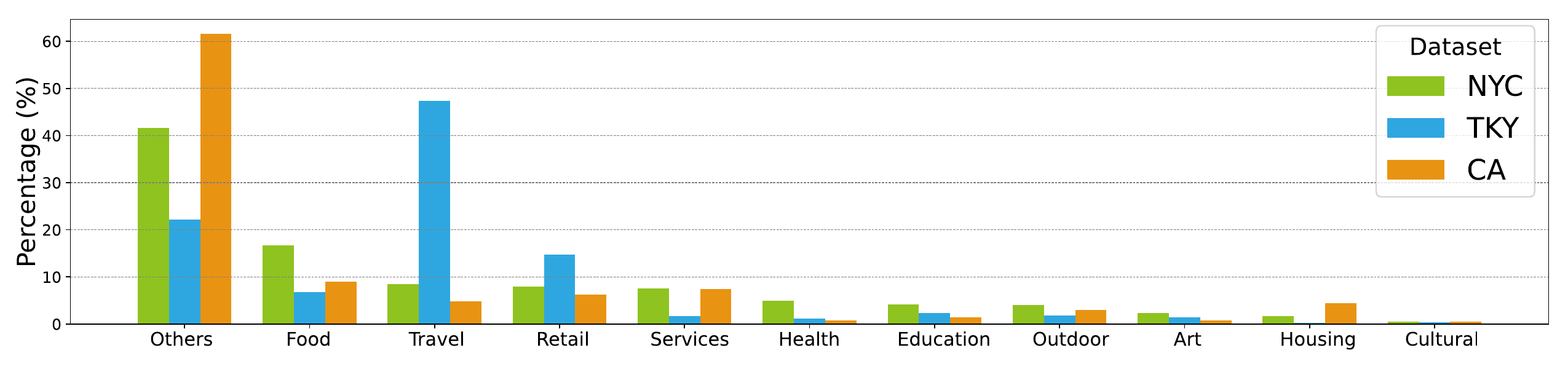}
    \caption{Statistics for different POI categories in the NYC, TKY, and CA datasets.}
    
  \label{fig:poi stats}
\end{figure*}

As shown in Table \ref{unseen}, interestingly, the models achieve competitive performance on datasets which they are not fine-tuned on. Specifically, the model trained in NYC has a top-1 accuracy lower than STHGCN for TKY and better than the state-of-the-art models in CA. The model trained on TKY performs even better in NYC than in NYC and is also better than the state-of-the-art models in CA. The model trained on CA is better than the state-of-the-art models in NYC and has a top-1 accuracy lower than STHGCN for TKY. This suggests that our models generalize well to unseen data. 

\begin{table}[t]
\caption{The models are LLM4POI trained only on one of the NYC, TKY, and CA datasets and evaluated on the rest.}
\label{unseen}
\centering
\begin{tabular}{lccc}
\toprule
Trained on  &  \multicolumn{1}{c}{NYC} & \multicolumn{1}{c}{TKY} &  \multicolumn{1}{c}{CA} \\ 
\cline{2-4} 
&                Acc@1     & Acc@1  & Acc@1\\ \hline
NYC                 & 0.3372   & 0.2594              & 0.1885\\
TKY                 &  0.3463     & 0.3035              & 0.1960  \\
CA                 & 0.3344     & 0.2600     & 0.2065  \\ 
\bottomrule
\end{tabular}
\end{table}

We look into the prompts to further investigate the reason for the generalization ability. As shown in Figure \ref{fig:portion}, we find that 75.8\%, 73.5\%, and 55.6\% of the prompts in test sets of NYC, TKY, and CA, respectively, have the answer \ac{POI} IDs appear within the questions. These portions are positively correlated to the overall model performance on each dataset. We also observe that the models trained on different datasets behave closely on each dataset, as shown in Figure \ref{fig:fourth}. Specifically, in their correct predictions, the portions of the prompts with answer \ac{POI} IDs that appear within the questions are almost identical. This suggests: 
\begin{enumerate*}[label=(\roman*)]
    \item the reason why our models have good performance is that our historical trajectory block combined with key-query similarity accurately captures the useful information from the current user's historical trajectories and other users' trajectories;
    \item the models learn to extract the correct \ac{POI} IDs from the prompts directly, which helps them to generalize to unseen data.
\end{enumerate*}

\subsubsection{Contextual Information Analysis}
What distinguishes our method from other works is that we use models embedded with commonsense knowledge to exploit contextual information. To evaluate how contextual information helps our model, we replace the \ac{POI} category names in the prompts with texts of the same lengths with no meaning to mask the contextual information. We analyze the model on the NYC, TKY, and CA datasets and compare the results with prompts with and without contextual information.

Table \ref{mask1} shows the overall results for the models trained on NYC, TKY, and CA and tested with different prompts. Model performance drops by a small margin when we remove the contextual information from \ac{POI} category names in NYC. For TKY and CA, the model performance decreases by 6.4\% and 6.2\%, respectively.

\begin{table}[t]
\caption{The results for LLM4POI tested with different prompts.}
\label{mask1}
\centering
\begin{tabular}{lccc}
\toprule
Prompts   &  \multicolumn{1}{c}{NYC} &  \multicolumn{1}{c}{TKY}&  \multicolumn{1}{c}{CA}\\ 
\cline{2-4} 
&   Acc@1&   Acc@1&   Acc@1     \\ \midrule
w/ context             & 0.3372 & 0.3035 &0.2065  \\
w/o context             &  0.3310  &0.2840& 0.1935  \\
 
\bottomrule
\end{tabular}
\end{table}

To further investigate the effectiveness of the contextual information, we evaluate the model on different levels of user activity with and without contextual information, similar to Section \ref{sec:cold_start}. 
From Table \ref{mask2}, we observe a significant drop in performance for inactive and normal users and an increase in performance for active users in TKY. The performance increases for inactive users and decreases for normal and active users in NYC and CA. The contextual information affects the model performance in two opposite directions for datasets in two different countries. Figure \ref{fig:poi stats} shows the stats for POI categories of NYC, TKY, and CA. We can see that the distributions of POI categories are indeed different between the two countries, where almost half of the POIs in TKY are for travel and transportation. Because POI contextual information differs from two types of datasets, the models behave differently from TKY to NYC and CA. This supports the models' understanding of the inherent meaning behind POI contextual information for the three datasets.

\begin{table}[t]
\caption{The results for LLM4POI tested with different prompts in terms of users with different levels of activity.}
\label{mask2}
\centering
\begin{tabular}{llccc}
\toprule
User groups & Prompts & \multicolumn{1}{c}{NYC}  & \multicolumn{1}{c}{TKY} & \multicolumn{1}{c}{CA} \\ \cline{3-5} 
                     &                & Acc@1& Acc@1& Acc@1     \\ \midrule
Inactive             &w/ context            & 0.3417&0.3478&  0.2132 \\
Normal               & w/ context            & 0.3841 &0.3516& 0.2057 \\
Very active          & w/ context            & 0.3088 &0.2727& 0.1920  \\ \midrule

Inactive             &w/o context          &  0.3493 &0.2751&0.2148 \\
Normal               & w/o context          & 0.3623 &0.2715& 0.1951\\
Very active          & w/o context         & 0.3025  &0.2884&0.1732  \\ \bottomrule
\end{tabular}
\end{table}

\subsubsection{Effect of Different Components}
We consider the performance of our model as the joint effect of 
\begin{enumerate*}[label=(\roman*)]
    \item the historical trajectory block;
    \item key-query similarity;
    \item contextual information.
\end{enumerate*}
To investigate the effect of each component, we remove the historical trajectory block in the prompt and only put the current user's historical trajectories in the historical trajectory block, respectively.

As shown in Table \ref{tab:abl}, the results suggest that each component contributes to the full model performance. Specifically, the historical trajectory block plays a critical role, where the top-1 accuracy drops as much as 50\% in TKY and CA. Because without any historical trajectories, the model suffers from short trajectories, which the datasets mostly consist of. With key-query similarity being removed, the collaborative information from other users is missing from the historical block, which leads to the inability to deal with the cold-start problem. On top of these two components, contextual information provides a further understanding of the data, improving the model's performance.

\begin{table}[ht]
\caption{Ablation study results for LLM4POI over three datasets.}
\label{tab:abl}
\centering
\begin{tabular}{lccc}
\toprule
Model & \multicolumn{1}{c}{NYC}  & \multicolumn{1}{c}{TKY} & \multicolumn{1}{c}{CA} \\ \cline{2-4} 
              & Acc@1& Acc@1& Acc@1     \\ \midrule
Full model          &  0.3372 &0.3035&0.2065 \\
w/o history          & 0.2356 &0.1517& 0.1016\\
w/o similarity         & 0.3171  &0.2836&0.1683  \\
w/o context             &  0.3310  &0.2840& 0.1935  \\\bottomrule
\end{tabular}
\end{table}

\section{Conclusion and Future Work}
In this paper, we propose LLM4POI, a framework to deploy \aclp{LLM} for the next \acl{POI} recommendation task, which is the first to utilize models with commonsense knowledge for the task. 
We developed trajectory prompting to transform the task into a question-answering. We also introduce key-query similarity to alleviate the cold-start problem. 

Our comprehensive experiments conducted on three real-world datasets show that we outperform all baseline models by a large margin. 
Our analysis supports that our method is able to handle the cold-start problem and various lengths of trajectories. 
It also shows the effectiveness of contextual information in our model. 
We have also shown the potential of developing foundation models for the next \ac{POI} recommendation task, given the ability of \aclp{LLM} to generalize to unseen data.

Because of the nature of \aclp{LLM}, we have limitations with efficiency regarding model training and inference time. Our design of the prompt is also limited by the context length and pretrained corpus of the model, e.g., the deprecation of geo-coordinates.  
For future work, we plan to address the limitations just mentioned. 
In addition, we will investigate chain-of-thought reasoning for the next \acl{POI} recommendation task, to both further boost the performance and provide explanations for the prediction. 
Another line of future work is to extend our models to scenarios without the extreme focus on a single best item, going beyond a question-answering context that we focused on here.

\subsubsection*{\bf Acknowledgements}
This research was conducted by the ARC Centre of Excellence for Automated Decision-Making and Society (CE200100005), and funded by the Australian Government through the Australian Research Council. In addition, this research was undertaken with the assistance of resources and services from the National Computational Infrastructure (NCI), which is supported by the Australian Government. Peibo Li is supported by Google Conference Scholarships (APAC).
Maarten de Rijke was partially supported by the Dutch Research Council (NWO), under project numbers 024.004.022, NWA.\-1389.20.183, and KICH3.LTP.20.006.

All content represents the opinion of the authors, which is not necessarily shared or endorsed by their respective employers and/or sponsors.

\balance
\normalem
\bibliographystyle{ACM-Reference-Format}
\bibliography{references}


\begin{thebibliography}{37}


\ifx \showCODEN    \undefined \def \showCODEN     #1{\unskip}     \fi
\ifx \showDOI      \undefined \def \showDOI       #1{#1}\fi
\ifx \showISBNx    \undefined \def \showISBNx     #1{\unskip}     \fi
\ifx \showISBNxiii \undefined \def \showISBNxiii  #1{\unskip}     \fi
\ifx \showISSN     \undefined \def \showISSN      #1{\unskip}     \fi
\ifx \showLCCN     \undefined \def \showLCCN      #1{\unskip}     \fi
\ifx \shownote     \undefined \def \shownote      #1{#1}          \fi
\ifx \showarticletitle \undefined \def \showarticletitle #1{#1}   \fi
\ifx \showURL      \undefined \def \showURL       {\relax}        \fi
\providecommand\bibfield[2]{#2}
\providecommand\bibinfo[2]{#2}
\providecommand\natexlab[1]{#1}
\providecommand\showeprint[2][]{arXiv:#2}

\bibitem[Brown et~al\mbox{.}(2020)]%
        {NEURIPS2020_1457c0d6}
\bibfield{author}{\bibinfo{person}{Tom Brown}, \bibinfo{person}{Benjamin Mann}, \bibinfo{person}{Nick Ryder}, \bibinfo{person}{Melanie Subbiah}, \bibinfo{person}{Jared~D Kaplan}, \bibinfo{person}{Prafulla Dhariwal}, \bibinfo{person}{Arvind Neelakantan}, \bibinfo{person}{Pranav Shyam}, \bibinfo{person}{Girish Sastry}, \bibinfo{person}{Amanda Askell}, {et~al\mbox{.}}} \bibinfo{year}{2020}\natexlab{}.
\newblock \showarticletitle{Language Models are Few-shot Learners}. In \bibinfo{booktitle}{\emph{Advances in neural information processing systems}}, Vol.~\bibinfo{volume}{33}. \bibinfo{pages}{1877--1901}.
\newblock


\bibitem[Chang et~al\mbox{.}(2023)]%
        {chang2023llm4ts}
\bibfield{author}{\bibinfo{person}{Ching Chang}, \bibinfo{person}{Wen-Chih Peng}, {and} \bibinfo{person}{Tien-Fu Chen}.} \bibinfo{year}{2023}\natexlab{}.
\newblock \showarticletitle{LLM4TS: Two-Stage Fine-Tuning for Time-Series Forecasting with Pre-Trained LLMs}. \bibinfo{howpublished}{arXiv preprint arXiv:2308.08469}.
\newblock


\bibitem[Chen et~al\mbox{.}(2023)]%
        {longlora}
\bibfield{author}{\bibinfo{person}{Yukang Chen}, \bibinfo{person}{Shengju Qian}, \bibinfo{person}{Haotian Tang}, \bibinfo{person}{Xin Lai}, \bibinfo{person}{Zhijian Liu}, \bibinfo{person}{Song Han}, {and} \bibinfo{person}{Jiaya Jia}.} \bibinfo{year}{2023}\natexlab{}.
\newblock \showarticletitle{LongLoRA: Efficient Fine-tuning of Long-Context Large Language Models}.
\newblock \bibinfo{journal}{\emph{arXiv:2309.12307}}.
\newblock


\bibitem[Cheng et~al\mbox{.}(2013)]%
        {cheng2013you}
\bibfield{author}{\bibinfo{person}{Chen Cheng}, \bibinfo{person}{Haiqin Yang}, \bibinfo{person}{Michael~R. Lyu}, {and} \bibinfo{person}{Irwin King}.} \bibinfo{year}{2013}\natexlab{}.
\newblock \showarticletitle{Where You Like to Go Next: Successive Point-of-Interest Recommendation}. In \bibinfo{booktitle}{\emph{{IJCAI} 2013, Proceedings of the 23rd International Joint Conference on Artificial Intelligence, Beijing, China, August 3-9, 2013}}, \bibfield{editor}{\bibinfo{person}{Francesca Rossi}} (Ed.). \bibinfo{publisher}{{IJCAI/AAAI}}, \bibinfo{pages}{2605--2611}.
\newblock


\bibitem[Cho et~al\mbox{.}(2011)]%
        {cho2011friendship}
\bibfield{author}{\bibinfo{person}{Eunjoon Cho}, \bibinfo{person}{Seth~A Myers}, {and} \bibinfo{person}{Jure Leskovec}.} \bibinfo{year}{2011}\natexlab{}.
\newblock \showarticletitle{Friendship and Mobility: User Movement in Location-based Social Networks}. In \bibinfo{booktitle}{\emph{Proceedings of the 17th ACM SIGKDD International Conference on Knowledge Discovery and Data Mining}}. \bibinfo{pages}{1082--1090}.
\newblock


\bibitem[Dao(2023)]%
        {dao2023flashattention}
\bibfield{author}{\bibinfo{person}{Tri Dao}.} \bibinfo{year}{2023}\natexlab{}.
\newblock \showarticletitle{FlashAttention-2: Faster Attention with Better Parallelism and Work Partitioning}.
\newblock \bibinfo{journal}{\emph{arXiv preprint arXiv:2307.08691}}.
\newblock


\bibitem[Dao et~al\mbox{.}(2022)]%
        {dao2022flashattention}
\bibfield{author}{\bibinfo{person}{Tri Dao}, \bibinfo{person}{Dan Fu}, \bibinfo{person}{Stefano Ermon}, \bibinfo{person}{Atri Rudra}, {and} \bibinfo{person}{Christopher R{\'e}}.} \bibinfo{year}{2022}\natexlab{}.
\newblock \showarticletitle{FlashAttention: Fast and Memory-efficient Exact Attention with IO-awareness}.
\newblock \bibinfo{journal}{\emph{Advances in Neural Information Processing Systems}}  \bibinfo{volume}{35} (\bibinfo{year}{2022}), \bibinfo{pages}{16344--16359}.
\newblock


\bibitem[Dettmers et~al\mbox{.}(2021)]%
        {dettmers20218}
\bibfield{author}{\bibinfo{person}{Tim Dettmers}, \bibinfo{person}{Mike Lewis}, \bibinfo{person}{Sam Shleifer}, {and} \bibinfo{person}{Luke Zettlemoyer}.} \bibinfo{year}{2021}\natexlab{}.
\newblock \showarticletitle{8-bit Optimizers via Block-wise Quantization}. In \bibinfo{booktitle}{\emph{International Conference on Learning Representations}}.
\newblock


\bibitem[Dettmers et~al\mbox{.}(2023)]%
        {dettmers2023qlora}
\bibfield{author}{\bibinfo{person}{Tim Dettmers}, \bibinfo{person}{Artidoro Pagnoni}, \bibinfo{person}{Ari Holtzman}, {and} \bibinfo{person}{Luke Zettlemoyer}.} \bibinfo{year}{2023}\natexlab{}.
\newblock \showarticletitle{QLoRA: Efficient Finetuning of Quantized LLMs}.
\newblock \bibinfo{journal}{\emph{arXiv preprint arXiv:2305.14314}}.
\newblock


\bibitem[Devlin et~al\mbox{.}(2019)]%
        {devlin-etal-2019-bert}
\bibfield{author}{\bibinfo{person}{Jacob Devlin}, \bibinfo{person}{Ming-Wei Chang}, \bibinfo{person}{Kenton Lee}, {and} \bibinfo{person}{Kristina Toutanova}.} \bibinfo{year}{2019}\natexlab{}.
\newblock \showarticletitle{{BERT}: Pre-training of Deep Bidirectional Transformers for Language Understanding}. In \bibinfo{booktitle}{\emph{Proceedings of the 2019 Conference of the North {A}merican Chapter of the Association for Computational Linguistics: Human Language Technologies, Volume 1 (Long and Short Papers)}}, \bibfield{editor}{\bibinfo{person}{Jill Burstein}, \bibinfo{person}{Christy Doran}, {and} \bibinfo{person}{Thamar Solorio}} (Eds.). \bibinfo{publisher}{Association for Computational Linguistics}, \bibinfo{address}{Minneapolis, Minnesota}, \bibinfo{pages}{4171--4186}.
\newblock
\urldef\tempurl%
\url{https://doi.org/10.18653/v1/N19-1423}
\showDOI{\tempurl}


\bibitem[Fairstein et~al\mbox{.}(2022)]%
        {farstein-2022-external}
\bibfield{author}{\bibinfo{person}{Yaron Fairstein}, \bibinfo{person}{Elad Haramaty}, \bibinfo{person}{Arnon Lazerson}, {and} \bibinfo{person}{Liane Lewin-Eytan}.} \bibinfo{year}{2022}\natexlab{}.
\newblock \showarticletitle{External Evaluation of Ranking Models under Extreme Position-Bias}. In \bibinfo{booktitle}{\emph{Proceedings of the Fifteenth ACM International Conference on Web Search and Data Mining}}. \bibinfo{publisher}{Association for Computing Machinery}, \bibinfo{address}{New York, NY, USA}, \bibinfo{pages}{252–261}.
\newblock


\bibitem[Feng et~al\mbox{.}(2015)]%
        {feng2015personalized}
\bibfield{author}{\bibinfo{person}{Shanshan Feng}, \bibinfo{person}{Xutao Li}, \bibinfo{person}{Yifeng Zeng}, \bibinfo{person}{Gao Cong}, {and} \bibinfo{person}{Yeow~Meng Chee}.} \bibinfo{year}{2015}\natexlab{}.
\newblock \showarticletitle{Personalized Ranking Metric Embedding for Next New POI Recommendation}. In \bibinfo{booktitle}{\emph{IJCAI'15 Proceedings of the 24th International Conference on Artificial Intelligence}}. ACM, \bibinfo{pages}{2069--2075}.
\newblock


\bibitem[Harte et~al\mbox{.}(2023)]%
        {harte2023leveraging}
\bibfield{author}{\bibinfo{person}{Jesse Harte}, \bibinfo{person}{Wouter Zorgdrager}, \bibinfo{person}{Panos Louridas}, \bibinfo{person}{Asterios Katsifodimos}, \bibinfo{person}{Dietmar Jannach}, {and} \bibinfo{person}{Marios Fragkoulis}.} \bibinfo{year}{2023}\natexlab{}.
\newblock \showarticletitle{Leveraging Large Language Models for Sequential Recommendation}. In \bibinfo{booktitle}{\emph{Proceedings of the 17th ACM Conference on Recommender Systems}}. \bibinfo{pages}{1096--1102}.
\newblock


\bibitem[He et~al\mbox{.}(2016)]%
        {he_inferring_2016}
\bibfield{author}{\bibinfo{person}{Jing He}, \bibinfo{person}{Xin Li}, \bibinfo{person}{Lejian Liao}, \bibinfo{person}{Dandan Song}, {and} \bibinfo{person}{William Cheung}.} \bibinfo{year}{2016}\natexlab{}.
\newblock \showarticletitle{Inferring a Personalized Next Point-of-interest Recommendation Model with Latent Behavior Patterns}. In \bibinfo{booktitle}{\emph{Proceedings of the {AAAI} {Conference} on {Artificial} {Intelligence}}}, Vol.~\bibinfo{volume}{30}.
\newblock
\newblock
\shownote{Issue: 1}.


\bibitem[Hochreiter and Schmidhuber(1997)]%
        {hochreiter1997long}
\bibfield{author}{\bibinfo{person}{Sepp Hochreiter} {and} \bibinfo{person}{J{\"u}rgen Schmidhuber}.} \bibinfo{year}{1997}\natexlab{}.
\newblock \showarticletitle{Long Short-term Memory}.
\newblock \bibinfo{journal}{\emph{Neural Computation}} \bibinfo{volume}{9}, \bibinfo{number}{8} (\bibinfo{year}{1997}), \bibinfo{pages}{1735--1780}.
\newblock


\bibitem[Hu et~al\mbox{.}(2021)]%
        {hu2021lora}
\bibfield{author}{\bibinfo{person}{Edward~J Hu}, \bibinfo{person}{Yelong Shen}, \bibinfo{person}{Phillip Wallis}, \bibinfo{person}{Zeyuan Allen-Zhu}, \bibinfo{person}{Yuanzhi Li}, \bibinfo{person}{Shean Wang}, \bibinfo{person}{Lu Wang}, {and} \bibinfo{person}{Weizhu Chen}.} \bibinfo{year}{2021}\natexlab{}.
\newblock \showarticletitle{LoRA: Low-rank Adaptation of Large Language Models}.
\newblock \bibinfo{journal}{\emph{arXiv preprint arXiv:2106.09685}}.
\newblock


\bibitem[Kong and Wu(2018)]%
        {kong_hst-lstm_2018}
\bibfield{author}{\bibinfo{person}{Dejiang Kong} {and} \bibinfo{person}{Fei Wu}.} \bibinfo{year}{2018}\natexlab{}.
\newblock \showarticletitle{{HST}-{LSTM}: {A} Hierarchical Spatial-Temporal Long-short Term Memory Network for Location Prediction.}. In \bibinfo{booktitle}{\emph{{IJCAI}}}, Vol.~\bibinfo{volume}{18}. \bibinfo{pages}{2341--2347}.
\newblock
\newblock
\shownote{Issue: 7}.


\bibitem[Lim et~al\mbox{.}(2020)]%
        {lim_stpudgat_2020}
\bibfield{author}{\bibinfo{person}{Nicholas Lim}, \bibinfo{person}{Bryan Hooi}, \bibinfo{person}{See-Kiong Ng}, \bibinfo{person}{Xueou Wang}, \bibinfo{person}{Yong~Liang Goh}, \bibinfo{person}{Renrong Weng}, {and} \bibinfo{person}{Jagannadan Varadarajan}.} \bibinfo{year}{2020}\natexlab{}.
\newblock \showarticletitle{{STP}-{UDGAT}: {Spatial}-Temporal-Preference User Dimensional Graph Attention Network for Next {POI} Recommendation}. In \bibinfo{booktitle}{\emph{Proceedings of the 29th {ACM} {International} {Conference} on {Information} \& {Knowledge} {Management}}}. \bibinfo{pages}{845--854}.
\newblock


\bibitem[Luo et~al\mbox{.}(2021)]%
        {luo2021stan}
\bibfield{author}{\bibinfo{person}{Yingtao Luo}, \bibinfo{person}{Qiang Liu}, {and} \bibinfo{person}{Zhaocheng Liu}.} \bibinfo{year}{2021}\natexlab{}.
\newblock \showarticletitle{Stan: Spatio-Temporal Attention Network for Next Location Recommendation}. In \bibinfo{booktitle}{\emph{Proceedings of the web conference 2021}}. \bibinfo{pages}{2177--2185}.
\newblock


\bibitem[Radford et~al\mbox{.}(2021)]%
        {radford2021learning}
\bibfield{author}{\bibinfo{person}{Alec Radford}, \bibinfo{person}{Jong~Wook Kim}, \bibinfo{person}{Chris Hallacy}, \bibinfo{person}{Aditya Ramesh}, \bibinfo{person}{Gabriel Goh}, \bibinfo{person}{Sandhini Agarwal}, \bibinfo{person}{Girish Sastry}, \bibinfo{person}{Amanda Askell}, \bibinfo{person}{Pamela Mishkin}, \bibinfo{person}{Jack Clark}, {et~al\mbox{.}}} \bibinfo{year}{2021}\natexlab{}.
\newblock \showarticletitle{Learning Transferable Visual Models from Natural Language Supervision}. In \bibinfo{booktitle}{\emph{International conference on machine learning}}. PMLR, \bibinfo{pages}{8748--8763}.
\newblock


\bibitem[Rendle et~al\mbox{.}(2010)]%
        {rendle2010factorizing}
\bibfield{author}{\bibinfo{person}{Steffen Rendle}, \bibinfo{person}{Christoph Freudenthaler}, {and} \bibinfo{person}{Lars Schmidt-Thieme}.} \bibinfo{year}{2010}\natexlab{}.
\newblock \showarticletitle{Factorizing Personalized Markov Chains for Next-basket Recommendation}. In \bibinfo{booktitle}{\emph{Proceedings of the 19th international conference on World wide web}}. \bibinfo{pages}{811--820}.
\newblock


\bibitem[Sun et~al\mbox{.}(2020)]%
        {sun_where_2020}
\bibfield{author}{\bibinfo{person}{Ke Sun}, \bibinfo{person}{Tieyun Qian}, \bibinfo{person}{Tong Chen}, \bibinfo{person}{Yile Liang}, \bibinfo{person}{Quoc Viet~Hung Nguyen}, {and} \bibinfo{person}{Hongzhi Yin}.} \bibinfo{year}{2020}\natexlab{}.
\newblock \showarticletitle{Where to Go Next: {Modeling} Long- and Short-term User Preferences for Point-of-interest Recommendation}. In \bibinfo{booktitle}{\emph{Proceedings of the {AAAI} {Conference} on {Artificial} {Intelligence}}}, Vol.~\bibinfo{volume}{34}. \bibinfo{pages}{214--221}.
\newblock
\newblock
\shownote{Issue: 01}.


\bibitem[Touvron et~al\mbox{.}(2023)]%
        {touvron2023llama}
\bibfield{author}{\bibinfo{person}{Hugo Touvron}, \bibinfo{person}{Louis Martin}, \bibinfo{person}{Kevin Stone}, \bibinfo{person}{Peter Albert}, \bibinfo{person}{Amjad Almahairi}, \bibinfo{person}{Yasmine Babaei}, \bibinfo{person}{Nikolay Bashlykov}, \bibinfo{person}{Soumya Batra}, \bibinfo{person}{Prajjwal Bhargava}, \bibinfo{person}{Shruti Bhosale}, {et~al\mbox{.}}} \bibinfo{year}{2023}\natexlab{}.
\newblock \showarticletitle{Llama 2: Open Foundation and Fine-tuned Chat Models}.
\newblock \bibinfo{journal}{\emph{arXiv preprint arXiv:2307.09288}}.
\newblock


\bibitem[Veli{\v{c}}kovi{\'c} et~al\mbox{.}(2018)]%
        {velivckovic2018graph}
\bibfield{author}{\bibinfo{person}{Petar Veli{\v{c}}kovi{\'c}}, \bibinfo{person}{Guillem Cucurull}, \bibinfo{person}{Arantxa Casanova}, \bibinfo{person}{Adriana Romero}, \bibinfo{person}{Pietro Li{\`o}}, {and} \bibinfo{person}{Yoshua Bengio}.} \bibinfo{year}{2018}\natexlab{}.
\newblock \showarticletitle{Graph Attention Networks}. In \bibinfo{booktitle}{\emph{International Conference on Learning Representations}}.
\newblock


\bibitem[Wang et~al\mbox{.}(2023)]%
        {wang2023would}
\bibfield{author}{\bibinfo{person}{Xinglei Wang}, \bibinfo{person}{Meng Fang}, \bibinfo{person}{Zichao Zeng}, {and} \bibinfo{person}{Tao Cheng}.} \bibinfo{year}{2023}\natexlab{}.
\newblock \showarticletitle{Where Would I Go Next? Large Language Models as Human Mobility Predictors}.
\newblock \bibinfo{journal}{\emph{arXiv preprint arXiv:2308.15197}} (\bibinfo{year}{2023}).
\newblock


\bibitem[Wang et~al\mbox{.}(2022)]%
        {wang2022learning}
\bibfield{author}{\bibinfo{person}{Zhaobo Wang}, \bibinfo{person}{Yanmin Zhu}, \bibinfo{person}{Haobing Liu}, {and} \bibinfo{person}{Chunyang Wang}.} \bibinfo{year}{2022}\natexlab{}.
\newblock \showarticletitle{Learning Graph-based Disentangled Representations for Next POI Recommendation}. In \bibinfo{booktitle}{\emph{Proceedings of the 45th International ACM SIGIR Conference on Research and Development in Information Retrieval}}. \bibinfo{pages}{1154--1163}.
\newblock


\bibitem[Wu et~al\mbox{.}(2020)]%
        {wu2020personalized}
\bibfield{author}{\bibinfo{person}{Yuxia Wu}, \bibinfo{person}{Ke Li}, \bibinfo{person}{Guoshuai Zhao}, {and} \bibinfo{person}{Xueming Qian}.} \bibinfo{year}{2020}\natexlab{}.
\newblock \showarticletitle{Personalized Long- and Short-term Preference Learning for Next POI Recommendation}.
\newblock \bibinfo{journal}{\emph{IEEE Transactions on Knowledge and Data Engineering}} \bibinfo{volume}{34}, \bibinfo{number}{4} (\bibinfo{year}{2020}), \bibinfo{pages}{1944--1957}.
\newblock


\bibitem[Xue et~al\mbox{.}(2022a)]%
        {xue2022translating}
\bibfield{author}{\bibinfo{person}{Hao Xue}, \bibinfo{person}{Flora~D. Salim}, \bibinfo{person}{Yongli Ren}, {and} \bibinfo{person}{Charles~L.A. Clarke}.} \bibinfo{year}{2022}\natexlab{a}.
\newblock \showarticletitle{Translating Human Mobility Forecasting through Natural Language Generation}. In \bibinfo{booktitle}{\emph{Proceedings of the Fifteenth ACM International Conference on Web Search and Data Mining}}. \bibinfo{pages}{1224--1233}.
\newblock


\bibitem[Xue et~al\mbox{.}(2022b)]%
        {xue2022leveraging}
\bibfield{author}{\bibinfo{person}{Hao Xue}, \bibinfo{person}{Bhanu~Prakash Voutharoja}, {and} \bibinfo{person}{Flora~D. Salim}.} \bibinfo{year}{2022}\natexlab{b}.
\newblock \showarticletitle{Leveraging Language Foundation models for Human Mobility Forecasting}. In \bibinfo{booktitle}{\emph{Proceedings of the 30th International Conference on Advances in Geographic Information Systems}}. \bibinfo{pages}{1--9}.
\newblock


\bibitem[Yan et~al\mbox{.}(2023)]%
        {yan2023spatio}
\bibfield{author}{\bibinfo{person}{Xiaodong Yan}, \bibinfo{person}{Tengwei Song}, \bibinfo{person}{Yifeng Jiao}, \bibinfo{person}{Jianshan He}, \bibinfo{person}{Jiaotuan Wang}, \bibinfo{person}{Ruopeng Li}, {and} \bibinfo{person}{Wei Chu}.} \bibinfo{year}{2023}\natexlab{}.
\newblock \showarticletitle{Spatio-Temporal Hypergraph Learning for Next POI Recommendation}. In \bibinfo{booktitle}{\emph{Proceedings of the 46th International ACM SIGIR Conference on Research and Development in Information Retrieval}}. \bibinfo{pages}{403--412}.
\newblock


\bibitem[Yang et~al\mbox{.}(2014)]%
        {yang2014modeling}
\bibfield{author}{\bibinfo{person}{Dingqi Yang}, \bibinfo{person}{Daqing Zhang}, \bibinfo{person}{Vincent~W Zheng}, {and} \bibinfo{person}{Zhiyong Yu}.} \bibinfo{year}{2014}\natexlab{}.
\newblock \showarticletitle{Modeling User Activity Preference by Leveraging User Spatial Temporal Characteristics in LBSNs}.
\newblock \bibinfo{journal}{\emph{IEEE Transactions on Systems, Man, and Cybernetics: Systems}} \bibinfo{volume}{45}, \bibinfo{number}{1} (\bibinfo{year}{2014}), \bibinfo{pages}{129--142}.
\newblock


\bibitem[Yang et~al\mbox{.}(2022)]%
        {yang2022getnext}
\bibfield{author}{\bibinfo{person}{Song Yang}, \bibinfo{person}{Jiamou Liu}, {and} \bibinfo{person}{Kaiqi Zhao}.} \bibinfo{year}{2022}\natexlab{}.
\newblock \showarticletitle{GETNext: Trajectory Flow Map Enhanced Transformer for Next POI Recommendation}. In \bibinfo{booktitle}{\emph{Proceedings of the 45th International ACM SIGIR Conference on Research and Development in Information Retrieval (SIGIR '22)}}. \bibinfo{pages}{1144--1153}.
\newblock


\bibitem[Zhang et~al\mbox{.}(2022)]%
        {zhang2022next}
\bibfield{author}{\bibinfo{person}{Lu Zhang}, \bibinfo{person}{Zhu Sun}, \bibinfo{person}{Ziqing Wu}, \bibinfo{person}{Jie Zhang}, \bibinfo{person}{Yew~Soon Ong}, {and} \bibinfo{person}{Xinghua Qu}.} \bibinfo{year}{2022}\natexlab{}.
\newblock \showarticletitle{Next Point-of-interest Recommendation with Inferring Multi-step Future Preferences}. In \bibinfo{booktitle}{\emph{Proceedings of the 31st International Joint Conference on Artificial Intelligence (IJCAI)}}. \bibinfo{pages}{3751--3757}.
\newblock


\bibitem[Zhang and Wang(2023)]%
        {zhang2023prompt}
\bibfield{author}{\bibinfo{person}{Zizhuo Zhang} {and} \bibinfo{person}{Bang Wang}.} \bibinfo{year}{2023}\natexlab{}.
\newblock \showarticletitle{Prompt Learning for News Recommendation}. In \bibinfo{booktitle}{\emph{Proceedings of the 46th International ACM SIGIR Conference on Research and Development in Information Retrieval}}. \bibinfo{pages}{227--237}.
\newblock


\bibitem[Zhao et~al\mbox{.}(2020)]%
        {zhao2020spatio}
\bibfield{author}{\bibinfo{person}{Pengpeng Zhao}, \bibinfo{person}{Anjing Luo}, \bibinfo{person}{Yanchi Liu}, \bibinfo{person}{Jiajie Xu}, \bibinfo{person}{Zhixu Li}, \bibinfo{person}{Fuzhen Zhuang}, \bibinfo{person}{Victor~S Sheng}, {and} \bibinfo{person}{Xiaofang Zhou}.} \bibinfo{year}{2020}\natexlab{}.
\newblock \showarticletitle{Where to Go Next: A Spatio-tTemporal Gated Network for Next POI Recommendation}.
\newblock \bibinfo{journal}{\emph{IEEE Transactions on Knowledge and Data Engineering}} \bibinfo{volume}{34}, \bibinfo{number}{5} (\bibinfo{year}{2020}), \bibinfo{pages}{2512--2524}.
\newblock


\bibitem[Zhao et~al\mbox{.}(2023)]%
        {zhao2023survey}
\bibfield{author}{\bibinfo{person}{Wayne~Xin Zhao}, \bibinfo{person}{Kun Zhou}, \bibinfo{person}{Junyi Li}, \bibinfo{person}{Tianyi Tang}, \bibinfo{person}{Xiaolei Wang}, \bibinfo{person}{Yupeng Hou}, \bibinfo{person}{Yingqian Min}, \bibinfo{person}{Beichen Zhang}, \bibinfo{person}{Junjie Zhang}, \bibinfo{person}{Zican Dong}, {et~al\mbox{.}}} \bibinfo{year}{2023}\natexlab{}.
\newblock \showarticletitle{A Survey of Large Language Models}.
\newblock \bibinfo{journal}{\emph{arXiv preprint arXiv:2303.18223}}.
\newblock


\bibitem[Zhou et~al\mbox{.}(2022)]%
        {zhou2022learning}
\bibfield{author}{\bibinfo{person}{Kaiyang Zhou}, \bibinfo{person}{Jingkang Yang}, \bibinfo{person}{Chen~Change Loy}, {and} \bibinfo{person}{Ziwei Liu}.} \bibinfo{year}{2022}\natexlab{}.
\newblock \showarticletitle{Learning to Prompt for Vision-Language Models}.
\newblock \bibinfo{journal}{\emph{International Journal of Computer Vision}} \bibinfo{volume}{130}, \bibinfo{number}{9} (\bibinfo{year}{2022}), \bibinfo{pages}{2337--2348}.
\newblock


\end{thebibliography}

\end{document}